# Prediction of many-electron wavefunctions using atomic potentials

Fariba Nazari* and Jerry L. Whitten

Department of Chemistry
North Carolina State University
Raleigh, NC 27695 USA

* Permanent address: *Department of Chemistry, Institute for Advanced Studies in Basic Sciences, Zanjan 45137-66731, Iran*

email: whitten@ncsu.edu, nazari@iasbs.ac.ir

**Abstract**

For a given many-electron molecule, it is possible to define a corresponding one-electron Schrödinger equation, using potentials derived from simple atomic densities, whose solution predicts fairly accurate molecular orbitals for single- and multi-determinant wavefunctions for the molecule. The energy is not predicted and must be evaluated by calculating Coulomb and exchange interactions over the predicted orbitals. Potentials are found by minimizing the energy of predicted wavefunctions. There exist slightly less accurate average potentials for first-row atoms that can be used without modification in different molecules. For a test set of molecules representing different bonding environments, these average potentials give wavefunctions with energies that deviate from exact self-consistent field or configuration interaction energies by less than 0.08 eV and 0.03 eV per bond or valence electron pair, respectively.

## I. Introduction

The prediction of wavefunctions by methods other than self-consistent field calculations has a long history in molecular and solid state systems. [1,2] Pauling's early work on chemical bonding can be implemented quantitatively by forming hybridized atomic orbitals and then constructing local bond orbitals. For minimal basis sets, such constructions involve only a few parameters. Since a single-determinant wavefunction is invariant to a linear transformation of its orbitals, the wavefunction constructed from non-orthogonal bond orbitals is equal to one formed from symmetrized and orthogonalized bond orbitals. The resulting molecular orbitals are often found to be in fairly good agreement with self-consistent field calculations using the same basis.[3] If the basis is not minimal, the procedure becomes less intuitive, but is still practical. Since evaluation of the energy requires the electron repulsion integrals, a hybridization-bond orbital construction offers no computational advantage if the problem can be solved directly by self-consistent field (SCF) procedures. However, local bonding constructions can be used to partition large systems into interacting components. Ruedenberg, Head-Gordon and coworkers and have discussed how localized orbitals can be used to construct and analyze molecular wavefunctions.[4-5] Methods have been developed to solve for localized orbitals directly, effective potentials have been developed for solid state systems, and other investigations have used bonding parameters obtained for molecular fragments or localized components to describe large systems.[5-17]

If the objective is simply to determine an initial field for an SCF calculation, there are many options. The simplest is to construct an approximate electron density by summing over atomic densities as demonstrated in early work[2] and more recently by many others.[16-17] One can also proceed more formally by making use of rigorous electrostatic error bounds derived from $(\rho(1) - \rho'(1) | r_{12}^{-1} | \rho(2) - \rho'(2)) \geq 0$ to obtain an approximation, $\rho'$, of an exact density, $\rho(1) = \sum_k \varphi_k(1)\varphi_k(1)$ where $\varphi_k$ denotes an accurate molecular orbital and $\rho'$ is a proposed approximation of arbitrary functional form.[18] Exchange can be added approximately. The Hohenberg-Kohn theorem guarantees the existence of one-electron potentials that predict the density, but these potentials and the corresponding Fock-operator or Kohn-Sham potentials involve the full electron density and require iterative constructions over electron repulsion contributions.

In the present work, we consider simple one-particle potentials and the question of whether universal potentials exist that can be transferred between systems to construct useful wavefunctions. The goal is to predict very accurate molecular orbitals for arbitrary systems. A one-particle Schrödinger equation is constructed for a given many-electron system such that its solution matches as closely as possible (to within a unitary transformation) the many-electron SCF solution using the same basis. Simple potentials are shown to exist that predict accurate solutions. Potentials are derived from optimized densities associated with atoms in a molecule or system. These optimized densities are not physical densities corresponding to occupied atomic orbitals, but rather special constructions that give optimum one-particle potentials. The existence of such potentials might appear to be in conflict with the complexity of the Fock operator that correctly determines the mixing of basis functions to form molecular orbitals. However, we show that is not the case, and that one-electron potentials exist that predict remarkably good molecular orbitals for single-determinant wavefunctions. The energy is not predicted and must be evaluated by calculating Coulomb and exchange interactions over the molecular orbitals. We also determine average potentials that can be used without change over a range of bonding environments with only a small decrease in accuracy. In some cases, these orbitals are shown to be accurate enough for direct use in configuration interaction calculations, by passing completely an SCF calculation. For a test set of molecules representing different bonding environments, the average potentials give wavefunctions with energies that deviate from exact SCF or configuration interaction (CI) energies by less than 0.08 eV and 0.03 eV per bond or valence electron pair, respectively. In the present work, potentials are determined in the context of many-electron theory, but the same argument can be applied to density functional calculations.

## II. Method

We begin by considering a molecule or other system described by the Schrödinger equation

$$H_{exact}\psi = (\sum_i [-\tfrac{1}{2}\nabla_i^2 - \sum_q \frac{Z_q}{r_{qi}}] + \sum_{i<j} r_{ij}^{-1})\psi = E_{exact}\psi \qquad (1)$$

with electrons and nuclei designated by $i$ and $q$, respectively, and associate with the system a modified Hamiltonian, $H^0$, that contains additional one-particle potentials for each nucleus, $v_{ki}$,

$$H^0\psi = (\sum_i h_i)\psi = (\sum_i [-\tfrac{1}{2}\nabla_i^2 + \sum_q (-\frac{Z_q}{r_{qi}} + v_{qi})])\psi = E\psi \qquad (2)$$

$$\text{where } \psi = (norm)\det(\chi_1(1)\chi_2(2)\chi_3(3)...) \quad \text{and} \quad \chi_i = \varphi_m(spin) \qquad (3)$$

Spatial orbitals are obtained by solving the one-electron problem $h_i\varphi_m(i) = \varepsilon_m\varphi_m(i)$.

Two variational solutions, $<\psi|H_{exact}|\psi> \geq E_{exact}$ are of interest for a given basis: a) the single determinant SCF solution and b) a configuration interaction solution involving a selected number of configurations. The objective of the present work is to find potentials $v_{qi}$ that produce orbitals $\varphi_m$ that match as closely as possible the variational solutions of a) or b). We first consider the SCF solution.

Potentials are assumed to derive from densities centered at nuclei where densities are expanded as a linear combination of normalized spherical Gaussian functions, $\rho_\beta = (\frac{\beta}{\pi})^{3/2}\exp(-\beta r^2)$. For a single component (at nucleus $q$), the repulsive potential acting on particle $i$ is

$$v_i = v(r_i) = \int |\vec{r}_i - \vec{r}|^{-1}\rho_\beta dv = 2\sqrt{\frac{\beta}{\pi}} r_i^{-1}\int_0^{r_i}\exp(-\beta r^2)dr \qquad (4)$$

The matrix element for the above density at (0,0,0) and two single s-type Gaussian functions $f_a$ and $f_b$, with exponents $a$ and $b$ and origins $(A_x,A_y,A_z)$ and $(B_x,B_y,B_z)$, respectively, as given by Boys,[19] is

$$<f_a(1)|v_1|f_b(1)> = 2\pi^{-1/2}<f_a|f_b>\frac{1}{\sqrt{R^2}}\int_0^{\sqrt{\gamma R^2}}\exp(-t^2)dt$$

where

$$\gamma = [\frac{1}{a+b} + \frac{1}{\beta}]^{-1} \qquad (5)$$

$R^2 = R_x^2 + R_y^2 + R_z^2$ and $R_x = (aA_x + bB_x)/(a+b)$, similarly for $y, z$.

Other integrals can be derived by differentiation with respect to parameters in the above expression.[19] In the present work, up to three such densities are allowed on a given nucleus, $\sum_{\beta} c_{\beta}\rho_{\beta}$. It is further assumed that the total density is normalized to the nuclear charge, $Z_q = \sum_{\beta} c_{\beta}$ such that the nuclear plus the repulsive potential is asymptotically zero for each nucleus. In the following, we refer to the added potential as a QC-potential and the solution of the resulting one-particle Schrödinger equation as the QC-potential method. It might be argued that severe restrictions such as spherical densities and neutral atoms make it unlikely that the resulting potentials would produce useful results in view of asymmetries in bonding and charge transfer effects. However, spherical densities combined with those on neighboring atoms introduce directional effects, and, varying exponents affects the polarity of bonds, i.e., smaller exponents decrease the shielding making a nucleus more attractive, and conversely.

In order to reach an acceptable level of accuracy, however, potentials must be carefully optimized, but once density parameters have been determined, applications are straightforward. We have optimized densities and corresponding potentials for individual molecules by the following scheme:

1) Initial exponents and coefficients are specified as parameters for the constituent atomic densities (e.g., for benzene, exponents and coefficients for C and H densities). Suppose the parameter values are $w_1, w_2, w_3 \ldots w_p$.

2) The resulting one-electron eigenvalue problem is solved to determine energies and coefficients of basis functions in molecular orbitals, $\{\varepsilon_m, \varphi_m\}$. The lowest energy N spin orbitals are occupied, (e.g., for benzene, 21 spatial molecular orbitals).

3) A single determinant wavefunction is constructed from the predicted orbitals and its energy is evaluated using the exact Hamiltonian, $H_{exact}$, a step that requires all electron repulsion integrals. The energy, $<\psi|H_{exact}|\psi>$, is a function of the parameters, $E(w_1, w_2, w_3 \ldots w_n)$.

4) Based on the value of $E$ and the current set of parameters, new parameters are selected and the process is repeated until $<\psi|H_{exact}|\psi>$ is minimized. The Nelder-Mead simplex procedure is a convenient way to accomplish this since the selection of new parameter values depends only on $E$ and the history of its variation with prior choices of parameters.[20]

The result of the optimization procedure is a set of density parameters, $\{c_a, \beta\}$, for each atom $k$ in the molecule being considered, $\rho_k = \sum_{\beta} c_{\beta}\rho_{\beta} = \sum_{\beta} c_{\beta}(\frac{\beta}{\pi})^{3/2} \exp(-\beta r_k^2)$. We emphasize that the QC-potentials do not enter into the many-electron Schrödinger equation. Once the one-particle Schrödinger equation is solved and orbitals are predicted, the potentials no longer appear in the formalism. The QC-densities do not resemble physical atomic densities derived from occupied atomic or molecular orbitals, but, instead, are special constructions that generate the optimum one-particle Schrodinger equation.

**III. Application to single-determinant SCF wavefunctions and configuration interaction**

Optimizations have been carried out for a set of molecules representing different types of bonding. Wavefunctions are predicted, energies are evaluated and results are compared with all-electron SCF calculations using the same basis. We refer to the latter canonical SCF solution as "exact" for the given basis. Each molecule is described by a double zeta basis of near Hartree-Fock atomic orbitals, contracted as 1s, 2s, and 2p orbitals plus additional two-term functions formed by taking the two longest exponent components of the 2s and 2p orbitals as a separate basis functions, d-functions are included in one system.[21] In the present study, all atoms with the same atomic number in the same molecule have the same potential (density) and no distinction is made between atoms in different bonding environments. Thus, the present results are an upper bound on the energy that would be improved if densities for an atom were allowed to vary within the molecule.

Before discussing the results, we explore the possibility of "average" potentials for H, C, N, O and F that could be used, without modification, for different molecules. To construct average potentials, we have considered molecules with complex bonding environments, rather than atoms or simple molecules, so that interactions are averaged *in situ* over different environments. The present calculations are based on $C_6H_6$, $N_2C_4H_4$, $H_2NCH_2$-COOH, $C_6H_5$-F, $C_6H_5$-COOH, and HFCO. The first four systems were treated successively, keeping atomic parameters determined for preceding molecules invariant. Parameters for O and F were then averaged with parameters from partial optimizations of the latter two molecules. Parameters are given in Table 1. If a larger data base becomes available, it would be useful to revisit the definition of average potentials. We summarize all calculations in Table 2 and 3, reporting energies for single-determinant and configuration interaction wavefunctions for the exact calculations, molecule optimized potentials and average potentials. In the CI calculations, Table 3, the SCF step is eliminated completely and molecular orbitals produced by solving the one-electron potential problem are used directly in the CI. In the molecule optimized calculations, densities for atoms in the individual molecules were optimized subject to the constraint that all atoms with the same atomic number in the same molecule have the same potential (density) and making no distinction between atoms in different bonding environments. As noted earlier, the QC-method only predicts orbitals: Coulomb and exchange integrals over molecular orbitals are required for the single-determinant energy and all two-electron integrals are needed for the CI calculations.

**Table 1.** Atomic densities used to define average potentials.
Individual densities are normalized, $(\frac{a}{\pi})^{3/2} \exp(-ar^2)$, and coefficients, $c_a$, sum to the nuclear charge (see text). A positive coefficient denotes repulsion.

**hydrogen**
| | | |
|---|---|---|
| exponents | 0.21861602E+00 | 0.10000000E+00 |
| coefficients | 0.17622709E+01 | -0.76227086E+00 |

**carbon**
| | | | |
|---|---|---|---|
| exponents | 0.53457450E+01 | 0.21475927E+00 | 0.10000000E+00 |
| coefficients | 0.27873507E+01 | 0.41082448E+01 | -0.89559557E+00 |

**nitrogen**
| | | | |
|---|---|---|---|
| exponents | 0.83384142E+01 | 0.10144018E+01 | 0.30150945E+00 |
| coefficients | 0.24905255E+01 | 0.14189014E+01 | 0.30905731E+01 |

**oxygen**
| | | | |
|---|---|---|---|
| exponents | 0.14794128E+01 | 0.11772868E+02 | 0.23466618E+00 |
| coefficients | 0.30500554E+01 | 0.22611025E+01 | 0.26888420E+01 |

**fluorine**
| | | | |
|---|---|---|---|
| exponents | 0.14000000E+01 | 0.15952835E+02 | 0.21555628E+00 |
| coefficients | 0.46969672E+01 | 0.22320562E+01 | 0.20709766E+01 |

**Table 2.** Total energies of selected molecules from exact SCF calculations and from wavefunctions predicted by the QC-potential method.[a] Energies are reported for calculations based on exact SCF, optimized potential and average potential calculations. The error per valence electron pair (bonds plus lone pairs excluding 1s electrons) is also given.

|  | **1-determinant energies** | | | | | |
|---|---|---|---|---|---|---|
|  | Exact SCF | Optimized Potentials | average potentials | Error exact-opt | error exact-avg | error per e-pair (avg pot) eV |
| **$C_6H_6$** | -230.6485 | -230.6398 | -230.6398 | -0.0088 | -0.0088 | 0.016 |
| **$C_4H_4N_2$** | -262.5793 | -262.5649 | -262.5611 | -0.0144 | -0.0183 | 0.033 |
| **$C_5H_5N$** | -246.6176 | -246.6053 | -246.6010 | -0.0123 | -0.0166 | 0.030 |
| **$C_2H_4$** | -78.0194 | -78.0175 | -78.0169 | -0.0019 | -0.0025 | 0.011 |
| **$CH_4$** | -40.1874 | -40.1870 | -40.1851 | -0.0004 | -0.0023 | 0.016 |
| **$C_2H_2$** | -76.8089 | -76.8073 | -76.8048 | -0.0017 | -0.0041 | 0.022 |
| **$H_2O$** | -76.0079 | -76.0070 | -76.0025 | -0.0008 | -0.0053 | 0.036 |
| **$H_2CO$** | -113.8287 | -113.8244 | -113.8119 | -0.0044 | -0.0168 | 0.076 |
| **$C_2F_2H_2$** | -275.6546 | -275.6440 | -275.6256 | -0.0107 | -0.0290 | 0.066 |
| **FHCO** | -212.6781 | -212.6667 | -212.6517 | -0.0114 | -0.0263 | 0.079 |
| **$NC_4H_5$** | -208.7742 | -208.7673 | -208.7403 | -0.0069 | -0.0339 | 0.071 |
| **$NC_4H_4$** | -208.1265 | -208.1159 | -208.0955 | -0.0106 | -0.0311 | 0.047 |
| **$C_6H_5$-F** | -329.5020 | -329.4857 | -329.4842 | -0.0163 | -0.0178 | 0.040 |
| **$C_6H_5$-$NH_2$** | -285.6597 | -285.6458 | -285.6383 | -0.0139 | -0.0214 | 0.032 |
| **$C_6H_5$-COOH** | -418.1783 | -418.1466 | -418.1287 | -0.0317 | -0.0496 | 0.058 |
| **$C_5H_5$-COOH** | -380.2957 | -380.2702 | -380.2495 | -0.0255 | -0.0462 | 0.060 |
| **$H_2NCH_2$-COOH** (glycine) | -282.7387 | -282.7225 | -282.6974 | -0.0162 | -0.0413 | 0.075 |
| C$_{24}$H$_{12}$ (graphene model) | -915.6445 | -915.6043 | -915.5978 | -0.0402 | -0.0467 | 0.024 |
| C$_{20}$N$_4$H$_{16}$ (chlorin) | -984.2629 | -984.1899 | -984.1607 | -0.0731 | -0.1023 | 0.048 |
| C$_4$H$_4$N$_2$ incl 3d | -262.6582 | -262.6322 | -262.6288 | -0.0260 | -0.0295 | 0.053 |

[a]Energies are in hartree atomic units unless specified otherwise, 1 a.u. = 27.21 eV,   1 eV=23.06 kcal/mol.

Energies from the QC-potential method in Table 2 are found to be in fairly good agreement with the exact SCF values for both the molecule optimized potentials and remarkably also for the average potentials. Deviations from exact energies are less than 0.08 eV per valence electron pair (bonds plus lone pairs excluding 1s electrons), for SCF calculations using average potentials. As the molecules increase in size, the deviations in total energy increase, but the errors per valance electron pair remain very small. Even for chlorin, which contains different nitrogen and carbon environments, the error is small. Chlorin, the graphene model, and pyrazine including 3d functions also have basis sets larger than the double zeta basis used for the other molecules and it is encouraging that the average potentials continue to give only small errors. We shall discuss in the Analysis Section why such simple spherical potentials, that do not resemble the potential of the Fock operator, can achieve such high accuracy.

CI energies from the QC-potential method are compared in Table 3 with exact CI values. In the calculations, configurations are generated by a hierarchical procedure[22,23] that includes single and double excitations from determinants that have a second order energy of interaction of $2 \times 10^{-6}$ hartrees with determinants in the expansion with coefficients greater than 0.02. The resulting expansions contain $10^4$ - $10^5$ determinants in the test set of molecules and thus the treatments are not near the full CI limit. Not surprisingly, the CI errors are greatly reduced compared to the single-determinant errors since the CI expansions recover part of the defect in molecular orbitals, and if carried out completely, all three CI calculations would give the same result. Also, the virtual orbitals from the potential method are often better for construction of excited configurations than the orbitals from the canonical SCF treatment. The latter orbitals correspond to negative ion states since they are determined in the fully occupied field of the ground state and thus generally are too diffuse spatially for optimum electron correlation. The fact that small residual errors exist means that defects in orbitals are not fully recovered; for the larger systems of model graphene and chlorin, there are only 52 electrons are in the active CI space and adjustments of the lower energy orbitals cannot occur in the present CI treatments.

**Table 3.** Total energies of selected molecules from exact CI calculations and from wavefunctions predicted by the QC-potential method.[a] CI calculations are performed directly using molecular orbitals predicted by the one-electron potential calculations. The error per valence electron pair (bonds plus lone pairs excluding 1s electrons) is also given.

|  | CI-energies | | | | | |
|---|---|---|---|---|---|---|
|  | Exact | Optimized Potential | average potential | Error exact-opt | error exact-avg | Error per e-pair (avg pot) eV |
| **$C_6H_6$** | -231.1422 | -231.1399 | -231.1399 | -0.0023 | -0.0023 | 0.004 |
| **$C_4H_4N_2$** | -263.0950 | -263.0903 | -263.0896 | -0.0047 | -0.0054 | 0.010 |
| **$C_5H_5N$** | -247.1239 | -247.1190 | -247.1174 | -0.0049 | -0.0065 | 0.012 |
| **$C_2H_4$** | -78.2246 | -78.2245 | -78.2244 | -0.0001 | -0.0002 | 0.001 |
| **$CH_4$** | -40.3031 | -40.3043 | -40.3039 | 0.0012 | 0.0008 | 0.000 |
| **$C_2H_2$** | -77.0108 | -77.0103 | -77.0094 | -0.0005 | -0.0014 | 0.007 |
| **$H_2O$** | -76.1376 | -76.1373 | -76.1376 | -0.0003 | 0.0000 | 0.000 |
| **$H_2CO$** | -114.0570 | -114.0571 | -114.0574 | 0.0001 | 0.0004 | 0.000 |
| **$C_2F_2H_2$** | -276.0858 | -276.0849 | -276.0792 | -0.0009 | -0.0066 | 0.014 |
| **FHCO** | -213.0186 | -213.0197 | -213.0196 | 0.0011 | 0.0009 | 0.000 |
| **$NC_4H_5$** | -209.1973 | -209.1944 | -209.1868 | -0.0029 | -0.0105 | 0.022 |
| **$NC_4H_4$** | -208.5428 | -208.5405 | -208.5361 | -0.0023 | -0.0067 | 0.015 |
| **$C_6H_5$-F** | -330.0932 | -330.0875 | -330.0869 | -0.0057 | -0.0063 | 0.011 |
| **$C_6H_5$-$NH_2$** | -286.2393 | -286.2320 | -286.2288 | -0.0073 | -0.0106 | 0.016 |
| **$C_6H_5$-COOH** | -418.9238 | -418.9048 | -418.9082 | -0.0190 | -0.0160 | 0.019 |
| **C₅H₅-COOH** | -381.0090 | -380.9931 | -380.9844 | -0.0159 | -0.0246 | 0.032 |
| **H₂NCH₂-COOH** (glycine) | -283.2245 | -283.2230 | -283.2163 | -0.0015 | -0.0078 | 0.015 |
| **C₂₄H₁₂** (graphene model) | -916.1005 | -916.0885 | -916.0849 | -0.0120 | -0.0156 | 0.008 |
| **C₂₀N₄H₁₆** (chlorin) | -984.5685 | -984.5397 | -984.5248 | -0.0288 | -0.0437 | 0.021 |
| **C₄H₄N₂** incl 3d | -263.2789 | -263.2624 | -263.2614 | -0.0165 | -0.0176 | 0.033 |

ªEnergies are in hartree atomic units unless specified otherwise, 1 a.u. = 27.21 eV

Comparing the errors for optimized and average potentials in the tables shows errors can be further reduced if potentials are optimized for individual molecules. This means there would be merit in constructing a library of potentials for functional groups or substructures that would be assembled to generate a solution for a larger system. Note, for example, the high accuracy of the graphene model treated using the benzene optimized potential. The results clearly show that the anisotropy or directional bonding in molecules is largely accounted for by potentials from neighboring nuclei and that mixings of different basis functions on the same nucleus as well as bond formation are well described by the simple potentials.

## IV. Analysis

As noted above, the QC potential method predicts the wavefunction, but not the energy. It is useful to consider a numerical example to compare solutions. In Table 4, the SCF eigenvalues of pyridine (exact calculation) are tabulated and compared with those of obtained by solving the one-electron potential problem. The table also includes eigenvalues obtained by diagonalizing the Fock matrix (including electron repulsion integrals) over these predicted orbitals with no mixing allowed between the occupied and virtual orbitals. The latter calculation is a unitary transformation of the occupied orbitals and virtual orbitals separately that does not change the energy of the wavefunction. Several points are important:

    1) The overlap of the predicted orbitals after the unitary transformation with those from the canonical SCF treatment is very high in every case. Thus, the exact and predicted wavefunctions are very similar as would be the electron density.

    2) The eigenvalues of the one-electron potential system are not at all close to the Fock eigenvalues, yet the predicted molecular orbitals agree well with the exact orbitals to within a unitary transformation. Thus, the variation in the difference,

$$<\psi|\sum_{i<j}^{N} r_{ij}^{-1}|\psi> - <\psi|\sum_{q}^{nuclei}\sum_{i}^{N} v_{qi}|\psi> \qquad (6)$$

in the neighborhood of the exact solution must be small. For the QC-potential method to predict

**Table 4.** Eigenvalues and total energies of pyridine from molecular orbitals predicted by optimized potentials and average potentials compared with values from an exact SCF calculation using the same basis. Rediagonalization values are from a diagonalization of the Fock matrix with no mixing between the occupied and virtual space. Also reported are eigenvalues directly from the one-electron atomic potential Schrödinger equation (in italics). The overlap of orbitals from the rediagonalized average potential calculation with those from the exact SCF calculation are listed. Energies are in hartrees.

| SCF exact | Optimized Potential | optimized potential rediag | average potential | average potential rediag | overlap <average\|exact> |
|---|---|---|---|---|---|
| 0.2250 | *0.1571* | 0.2243 | *0.2592* | 0.2202 | 0.9940 |
| 0.1871 | *0.1211* | 0.1855 | *0.2279* | 0.1777 | 0.9923 |
| 0.1289 | *0.0673* | 0.1287 | *0.1938* | 0.1203 | 0.9949 |
| 0.1184 | *0.0537* | 0.1151 | *0.1785* | 0.1078 | 0.9996 |
| -0.3655 | *-0.1341* | -0.3663 | *-0.0154* | -0.3752 | 0.9984 |
| -0.3945 | *-0.1405* | -0.3965 | *-0.0228* | -0.3934 | 0.9996 |
| -0.4187 | *-0.1661* | -0.4089 | *-0.0400* | -0.4001 | 0.9958 |
| -0.5282 | *-0.2474* | -0.5240 | *-0.1286* | -0.5284 | 0.9959 |
| -0.5517 | *-0.2667* | -0.5536 | *-0.1435* | -0.5565 | 0.9919 |
| -0.5890 | *-0.2880* | -0.5852 | *-0.1656* | -0.5955 | 0.9965 |
| -0.6168 | *-0.3147* | -0.6146 | *-0.1942* | -0.6230 | 0.9919 |
| -0.6647 | *-0.3510* | -0.6635 | *-0.2286* | -0.6710 | 0.9949 |
| -0.6712 | *-0.3644* | -0.6709 | *-0.2476* | -0.6777 | 0.9953 |
| -0.7378 | *-0.4115* | -0.7379 | *-0.2886* | -0.7471 | 0.9942 |
| -0.8681 | *-0.5040* | -0.8678 | *-0.3854* | -0.8719 | 0.9940 |
| -0.8738 | *-0.5090* | -0.8746 | *-0.3918* | -0.8864 | 0.9959 |
| -1.0535 | *-0.6332* | -1.0540 | *-0.5178* | -1.0641 | 0.9984 |
| -1.1096 | *-0.6775* | -1.1093 | *-0.5588* | -1.1249 | 0.9982 |
| -1.2852 | *-0.8150* | -1.2845 | *-0.6996* | -1.2776 | 0.9948 |
| -11.2656 | *-10.0227* | -11.2575 | *-9.8727* | -11.2850 | 0.9954 |
| -11.2657 | *-10.0227* | -11.2576 | *-9.8748* | -11.2851 | 0.9993 |
| -11.2792 | *-10.0260* | -11.2775 | *-9.8749* | -11.3142 | 1.0000 |
| -11.2944 | *-10.0322* | -11.3015 | *-9.8862* | -11.3142 | 0.9993 |
| -11.2944 | *-10.0322* | -11.3016 | *-9.8862* | -11.3209 | 1.0000 |
| -15.5841 | *-14.1907* | -15.5828 | *-14.0722* | -15.5334 | 1.0000 |

**Sum of eigenvalues (occupied)**
| -82.7119 | -70.1497 | -82.6901 | -67.4953 | -82.8452 | |

**Total energy**
| -246.6176 | | -246.6053 | | -246.6010 | |

molecular orbitals accurately, it is only necessary that the mixing between orthonormal basis functions be correctly described and that mixing depends on ratios of matrix elements. Consider, for example, one such set of orthonormal basis functions - the exact SCF solution. If this basis of molecular orbitals is used to solve the one-electron potential problem, the solution would be exact only if all matrix elements between the occupied and virtual orbitals are zero, $H_{ij}^0 = <\varphi_i|H^0|\varphi_j> = 0$. A numerical example from $H_2CO$ provides a useful illustration. Molecular orbitals are obtained from the exact SCF solution for $H_2CO$ and a Hamiltonian, $H^0$, is defined by choosing the average potentials for C, O and H from Table 1. Matrix elements are calculated and in Table 5 values are listed for the second order energy contribution for each pair of occupied and virtual orbitals for which the matrix element is nonzero,

$$\Delta_{ij} = -(H_{ij}^0)^2 (H_{jj}^0 - H_{ii}^0)^{-1} \tag{7}$$

where $i$ and $j$ denote occupied and virtual orbitals, respectively. The quantity $\Delta_{ij}$ is a measure of the energy driving force toward an incorrect solution, i.e. a mixing of the occupied and virtual spaces by $H^0$. The value of $\Delta_{ij}$ for each pair of orbitals is found to be small. The sum is -0.0116 where a factor of two is included to account for double occupancy. This value is comparable to the actual calculated error of -0.0168 reported in Table 2.

We now examine contributions to the error more closely by a series of calculations on pyridine; calculations are for molecule optimized potentials:

|  | Double-zeta basis | Triple-zeta basis |
|---|---|---|
| Exact SCF | -246.6176 | -246.6513 |
| C, N  (3 component densities) | -246.6053 (0.0123) | -246.6312 (0.0201) |
| C, N  (4 component densities) | -246.6058 (0.0118) | -246.6325 (0.0188) |
| C, N  (1s constrained invariant) | -246.6089 (0.0087) | -246.6371 (0.0142) |

In general, the error, in parentheses, increases with the number of degrees of freedom as is evident comparing the double- and triple-zeta calculations. A major contribution to the error is due to a slightly incorrect mixing with the 1s orbitals suggesting that it would be better to include these orbitals in an invariant core. Doing so reduces the need for the potential to account for the short range mixing of the 1s with other functions. The calculations show no significant improvement in using four component densities. Preliminary studies on pyrazine also show a non-spherical potential for N atoms with nonbonded electrons will reduce the error.

The triple-zeta energies for the test set of molecules are given in Table 6 for the QC potentials determined for the double-zeta basis, i.e., those defined in Table 1. An increase in error occurs due to the extra degrees of freedom in the larger basis. However, for molecules containing C, N, and H, the errors per valence electron pair remain fairly small in the range 0.02 to 0.08 eV. The errors for molecules containing O and F are larger and suggest a need for re-optimization of the O and F potentials. Optimizations of O and F potentials were carried out for the triple-zeta basis and the reduced errors are also reported in Table 6. Further optimization of the C, N and H potentials to determine a new set of average potentials for triple-zeta type basis sets would be expected to reduce the error further. However, even without further optimization, the errors are comparable to those obtained with the double-zeta basis.

**Table 5.** Energy error due to mixing of the correct occupied orbitals with virtual orbitals in $H_2CO$ caused by a slightly incorrect average potential. If the potential were exact, the matrix elements $H_{ij}^0 = <\varphi_i | H_0 | \varphi_j >$ between occupied and virtual elements would be zero. Values are tabulated for the second order energy, $\Delta_{ij} = -(H_{ij}^0)^2 (H_{jj}^0 - H_{ii}^0)^{-1}$, for each pair of occupied, $i$, and virtual, $j$, orbitals where the molecular orbitals $\varphi_i$ are from an exact SCF calculation. The sum of energies ( ×2) is comparable to the actual error reported in Table 2 for the average potential. Energies are in hartrees.

| $i$ | $j$ | $\Delta_{ij}$ | $i$ | $j$ | $\Delta_{ij}$ |
|---|---|---|---|---|---|
| 1 | 12 | -2.06E-005 | 5 | 11 | -2.44E-005 |
| 1 | 13 | -7.61E-007 | 5 | 14 | -4.54E-004 |
| 1 | 16 | -2.76E-006 | 5 | 17 | -1.59E-004 |
| 1 | 18 | -9.38E-008 | 5 | 21 | -3.55E-004 |
| 1 | 19 | -1.54E-007 | | | |
| 1 | 22 | -3.45E-006 | 6 | 10 | -3.60E-005 |
| | | | 6 | 12 | -4.63E-004 |
| 2 | 10 | -9.89E-006 | 6 | 13 | -2.40E-006 |
| 2 | 12 | -4.56E-006 | 6 | 16 | -6.69E-005 |
| 2 | 13 | -3.91E-006 | 6 | 18 | -7.36E-007 |
| 2 | 16 | -1.10E-005 | 6 | 19 | -6.18E-004 |
| 2 | 18 | -5.52E-005 | 6 | 22 | -1.78E-006 |
| 2 | 19 | -5.73E-006 | | | |
| 2 | 22 | -5.46E-006 | 7 | 9 | -1.88E-003 |
| | | | 7 | 15 | -4.03E-006 |
| 3 | 10 | -3.86E-006 | | | |
| 3 | 12 | -3.96E-006 | 8 | 11 | -9.83E-005 |
| 3 | 13 | -1.16E-006 | 8 | 14 | -2.57E-004 |
| 3 | 16 | -4.85E-005 | 8 | 17 | -4.87E-005 |
| 3 | 18 | -2.06E-005 | 8 | 21 | -3.98E-004 |
| 3 | 19 | -2.06E-004 | | | |
| 3 | 22 | -6.36E-005 | | | |
| | | | Total (x2) | | -1.16E-002 |
| 4 | 12 | -1.92E-004 | | | |

| 4 | 16 | -3.50E-005 |
| 4 | 18 | -2.81E-006 |
| 4 | 19 | -2.18E-004 |
| 4 | 22 | -1.90E-005 |

**Table 6.** Single-determinant energies obtained using a triple-zeta valence basis. Energies are given for the exact scf calculation, and the energy of the wavefunction determined using the qc densities in Table 1. Energy per valence electron pair are given for the original potential and for re-opotimized O and F potentials, in brackets. Calculations are for all electrons, however, the 1s orbitals are constrained to be invariant in the qc potential calculations. Total energies are in hartrees.

|  | Exact SCF | Average | Error eV/e-pair |
|---|---|---|---|
| **$C_4H_4N_2$** | -262.6226 | -262.5943 | 0.051 |
| **$C_5H_5N$** | -246.6513 | -246.6298 | 0.039 |
| **$C_6H_6$** | -230.6728 | -230.6611 | 0.021 |
| **$H_2NCH_2$-COOH** (glycine) | -282.7876 | -282.7372 | 0.092 [0.086] |
| **$C_2H_2$** | -76.8119 | -76.8063 | 0.031 |
| **$C_5H_5$-COOH** | -380.3501 | -380.2947 | 0.072 [0.068] |
| **$C_6H_5$-COOH** | -418.2369 | -418.1774 | 0.070 [0.065] |
| **$C_6H_5$-$NH_2$** | -285.6964 | -285.6707 | 0.039 |
| **$CH_4$** | -40.1895 | -40.1867 | 0.019 |
| **$C_2H_4$** | -78.0241 | -78.0201 | 0.018 |
| **$H_2CO$** | -113.8444 | -113.8229 | 0.098 [0.090] |
| **$H_2O$** | -76.0185 | -76.0119 | 0.045 [0.044] |
| **$NC_4H_4$** | -208.1259 | -208.0893 | 0.077 |
| **$NC_4H_5$** | -208.7715 | -208.7398 | 0.066 |
| **$C_2F_2H_2$** | -275.6796 | -275.6024 | 0.175 [0.077] |
| **$C_6H_5$-F** | -329.5403 | -329.5064 | 0.051 [0.053] |
| **FHCO** | -212.7036 | -212.6525 | 0.154 [0.097] |

## V. Conclusions

For a given many-electron molecule, it is possible to define a corresponding one-electron Schrödinger equation, using potentials derived from simple atomic densities, whose solution predicts fairly accurate molecular orbitals for single- and multi-determinant wavefunctions for the molecule. The energy is not predicted and must be evaluated by calculating Coulomb and exchange interactions over the predicted orbitals. Potentials are found by minimizing the energy of predicted wavefunctions.

There exist slightly less accurate average potentials for first-row atoms that can be used without modification in different molecules. For a test set of molecules representing different bonding environments, these average potentials give wavefunctions with energies that deviate from exact SCF or CI energies by less than 0.08 eV and 0.03 eV per bond or valence electron pair, respectively. In the CI calculations, the SCF step is bypassed completely. The present work demonstrates that it is much easier to find simple potentials that predict an accurate wavefunction than simple potentials that would give an accurate energy.

The QC-average potentials produce excellent initial fields for canonical SCF solutions or orbitals for post-Hartree Fock methods that include correlation contributions. A library of densities/potentials optimized for atoms in different environments or for molecular substructures would further improve the accuracy of applications.

The one-electron potentials should be helpful in defining embedding potentials for treatment of a local region of a large system, e.g., a surface-adsorbate system, or a very large molecule.

The present study was carried out in the context of many-electron theory. The same arguments could be applied to density functional applications.

## VI. Supplementary Material

Coordinates for all systems calculated are provided in Supplementary Materials.

## Acknowledgment

This work originated in an advanced quantum mechanics class taught by one of the authors (JLW). The stimulating questions by students, particularly M.C. Bennett and C.A. Melton, is gratefully acknowledged as are helpful discussions with Professor Mike Whangbo.

1818